\documentclass[twocolumn]{jpsj3} 
\usepackage{txfonts}
\usepackage{graphicx,amsmath,xspace}
\usepackage{color}

\title{
  {Theory of Spin Motive Force} \\
{in One-Dimensional Antiferromagnetic Domain Wall} 
}

\author{
Akira Okabayashi
\thanks{E-mail: okabayashi.akira.73c@st.kyoto-u.ac.jp}
and
Takao Morinari
\thanks{E-mail: morinari.takao.5s@kyoto-u.ac.jp}
}

\inst{
Graduate School of Human and Environmental Studies, Kyoto University, 
Kyoto 606-8501, Japan
}

\newcommand{\tm}[1]{\textcolor{black}{#1}}

\newcommand{\be}{\begin{equation}}
\newcommand{\ee}{\end{equation}}
\newcommand{\bea}{\begin{eqnarray}}
\newcommand{\eea}{\end{eqnarray}}
\newcommand{\bsigma}{\boldsymbol{\sigma}}

\date{\today}

\abst{
We present the theory of the spin motive force in antiferromagnets.
We consider a one-dimensional antiferromagnetic domain wall
strongly coupled with conduction electrons via an exchange interaction.
We carry out a unitary transformation 
that rotates the spin coordinate system of the conduction electron locally,
so that the quantization axis is in the direction of the localized spin.
By numerically solving the time dependent 
Schr\"{o}dinger equation, we clearly demonstrate 
that the spin motive force acts on the conduction electron.
The result suggests that there is no distinction between
antiferromagnets and ferromagnets 
from the viewpoint of the basic phenomenon relevant to spintronics.
}

\begin{document}
\maketitle

In the field of spintronics,\cite{Wolf2001} 
one of the technical issues is 
creating spin-polarized current.
A metal with localized spins forming a spin texture,
such as a domain wall, 
can create a spin current under a magnetic field
in the presence of strong coupling 
between conduction electrons and localized spins.\cite{Berger1986}
The precession of localized spins induced by the magnetic field
leads to a time dependent Berry phase effect.\cite{Berry1984}
This Berry phase effect gives rise to a motive force,
\cite{Volovik1987,Stern1992,Barnes2007,Saslow2007,Duine2008,Tserkovnyak2008,Ohe2009}
which is called the spin motive force,
and creates spin-polarized current. 
Electrical voltage generated by such a Berry phase effect
has been confirmed experimentally.\cite{Hai2009,Yang2009,Yamane2011}
On the other hand, owing to the conservation of spin angular momentum,
a torque on localized spins,
the so-called spin-transfer torque effect,\cite{Ralph2008,Parkin2008}
is created by the spin-polarized current.


Most spintronics research focuses on ferromagnets,
but antiferromagnets can also be used to manipulate spin information.
\cite{MacDonald2011}
%
The important difference between ferromagnets and 
antiferromagnets is that 
the smoothly varying field is not local magnetization
but local staggered magnetization.
%
By using the smoothly varying 
local staggered magnetization\cite{Cheng2012,Tveten2013,Cheng2014}, 
current-induced torque effects in antiferromagnets 
have been predicted theoretically\tm{\cite{Nunez2006,Hals2011}}
and confirmed experimentally\cite{Wei2007,Jaramillo2007,Urazhdin2007} 
The reverse of this effect,
that is, the spin motive force for antiferromagnets,
has been formulated 
by a semiclassical approximation.\cite{Cheng2012,Cheng2014}
Starting from the band picture,
the spatial variation of spins is included within 
the semiclassical approximation.
An important issue is whether the semiclassical approximation
is justified in the case of the antiferromagnets.

In this work, we develop a theory for  
the spin motive force created by the staggered magnetization dynamics
without relying on the semiclassical approximation.
We consider a system consisting of 
a one-dimensional antiferromagnetic domain wall
and a conduction electron
with the exchange interaction
between localized spins and the conduction electron.
By solving the time dependent Schr\"{o}dinger equation
for a single conduction electron,
we show that, under a magnetic field along the system,
the rigidly precessing domain wall gives rise to the spin motive force
in the strong coupling limit of the exchange interaction.
This spin motive force is not described 
by the semiclassical approximation.
There is a combined effect of the sign change of the potential
created by the time dependent Berry phase and 
the spin flip at each hopping of the conduction electron.
This combined effect leads to the net spatial gradient
of the effective potential associated with the staggered potential,
and so the spin motive force acts on the conduction electron.

\begin{figure}
   \begin{center}
    \includegraphics[width=0.8 \linewidth]{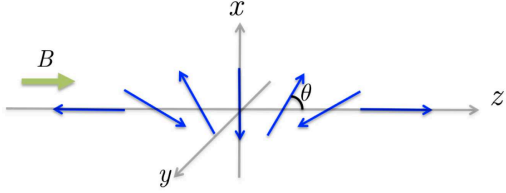}
   \end{center}
   \caption{ \label{fig_afdw}
     (Color online)
     Antiferromagnetic domain wall. 
     Arrows represent localized moments constituting
     the domain wall.
     We take the coordinate system as shown in the figure.
     The thick arrow represents the applied magnetic field.
     Under the magnetic field, the domain wall precesses
     about the $z$-axis.
     $\theta$ is the angle between the direction of the local moment
     and the $z$-axis.
     We define $\theta$ at each site and denote it by $\theta_j$
     at site $j$.
    }
 \end{figure}

We consider a one-dimensional antiferromagnet.
The coordinate axes are defined as shown in Fig.~\ref{fig_afdw}.
The localized moment with spin $S$ at site $j$ is represented by
\[
{{\bf{S}}_j} 
= S(\sin {\theta _j}\cos {\phi _j},\sin {\theta _j}
\sin {\phi _j},\cos {\theta _j}).
\]
We assume that there is a static antiferromagnetic domain wall,
\cite{Papanicolaou1995,Bode2006,Cheng2014} with
\[
{\theta _j} = \frac{\pi }{2}\left( {1 - \frac{j}{{{\ell}}}} \right),
\]
as shown in Fig.~\ref{fig_afdw}.
Here, $\ell$ is the length scale of the domain wall
and we take the lattice constant to be unity.

A conduction electron at site $j$
interacts with the localized spin at the same site
via the exchange interaction $J$.
We apply a magnetic field along the $z$-axis, ${\bf B}=(0,0,B)$.
Under this magnetic field, localized moments precess
about the $z$-axis.
Here, we assume that the domain wall is rigidly precessing
so that 
\[
\phi_j = 2\pi t/T.
\]
Here, $1/T \equiv g \mu _B B/(2\pi)$ with $g$ as the g-factor
and $\mu_B$ as the Bohr magneton.
To focus on the spin motive force,
we do not consider the spin transfer torque created by
the current.

The Hamiltonian of the system is given by
\be
{\cal H} =  - \eta \sum\limits_j {({c_j}^\dag {c_{j + 1}} 
  + {\rm{h}}.{\rm{c}}.)}  - J\sum\limits_j {{{\bf{S}}_j} 
  \cdot ({c_j}^\dag \bsigma {c_j})}.
\label{eq_H0}
\ee
Here, $\eta$ is the hopping parameter for the conduction electron.
The operator $c_j^\dag$ is a creation operator
of the conduction electron at site $j$ in the spinor form,
$c_j^\dag  = \left( {\begin{array}{*{20}{c}}
{c_{j \uparrow }^\dag }&{c_{j \downarrow }^\dag }
\end{array}} \right)$.
The vector $\bsigma$ has components of the Pauli matrices:
\[
\bsigma  = \left( {{\sigma _x},{\sigma _y},{\sigma _z}} \right).
\]
Note that 
the Hamiltonian does not contain the dynamics of the domain wall
because it is completely determined as mentioned above.
The time dependence of the localized moments 
enter the Hamiltonian through the time dependence of ${\bf S}_j$.

Now, we take the strong coupling limit with respect to $J$.
In the strong coupling limit, the spin of the conduction electron
at site $j$ is in the direction of ${\bf S}_j$.
We rotate the spin coordinate system of the conduction electron locally,
so that the $z$-direction is in the direction of ${\bf S}_j$.\cite{Prange1977}
Such a rotation is carried out by the following unitary transformation, 
\[
  {c_j} \to {U_j}{c_j},
  \]
with
\[
  {U_j} = {{\bf{d}}_j} \cdot \bsigma,
  \]
Here,
\[
  {{\bf{d}}_j} = \frac{1}{{\sqrt {2(1 + {S_{jz}})} }}
  \left( {{S_{jx}},{S_{jy}},1 + {S_{jz}}} \right).
  \]
After carrying out this transformation, the staggered potential
$(-1)^j JS\sigma_z$ acts on the conduction electron.
This potential term is $JS\sigma_z$ for $j$ even sites,
i.e., sublattice A,
and $-JS\sigma_z$ for $j$ odd sites, i.e., sublattice B.
To remove the sign difference in this potential term,
we perform an additional unitary transformation,
$c_j \rightarrow i\sigma_y c_j$, at sublattice B.
Thus, the Hamiltonian Eq. (\ref{eq_H0}) is rewritten as
\begin{align}
{\cal H} &=  - \eta \sum\limits_{j \in A} 
{\left( {{c_j}^\dag U_j^\dag {U_{j + 1}}i{\sigma _y}{c_{j + 1}} 
    + {c_j}^\dag U_j^\dag {U_{j - 1}}i{\sigma _y}{c_{j - 1}} + {\rm{h}}.{\rm{c}}.} \right)} \nonumber \\
 &  - JS\sum\limits_j {{c_j}^\dag {\sigma _z}{c_j}}  
+ i\hbar \sum\limits_{j \in A} 
{c_j^\dag \left( {U_j^\dag {\partial _t}{U_j}} \right){c_j}}  \nonumber \\
& - i\hbar \sum\limits_{j \in B} {c_j^\dag 
  \left( {U_j^\dag {\partial _t}{U_j}} \right){c_j}}.
\label{eq_H}
\end{align}
The third and fourth terms have been added so that
the Hamiltonian is consistent with the equation of motion of
the creation and annihilation operators.

\begin{figure}
    \begin{center}
      \includegraphics[width=0.8 \linewidth]{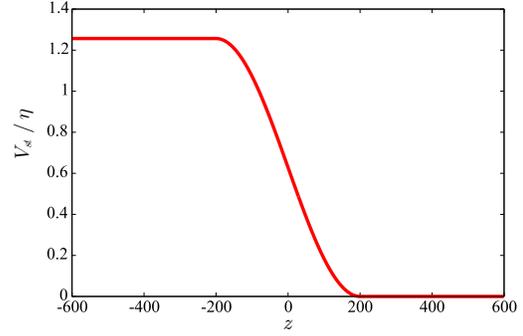}
    \end{center}
    \caption{ \label{fig_Vst}
     (Color online)
      Staggered potential $V_{st}$ as a function of $z$.
      Because of the antiferromagnetic nature,
      the time dependent Berry phase creates 
      the staggered potential $V_{st}$
      Eq.~(\ref{eq_Vst}), as described in the text.
    }
  \end{figure}

The effect of the domain wall dynamics on the conduction electron
is conveniently represented by the gauge field
\be
{a_{0j}} =  - i\hbar {U_j}^\dag {\partial _t}{U_j},
\ee
\be
{{\bf{a}}_j} = i\hbar {U_j}^\dag \tm{\nabla_j} {U_j}.\label{vec_po}
\ee
\tm{Here, we define $\nabla_j$ as $\nabla_jf_j=(f_{j+1}-f_{j-1})/2$ for the function $f_j$.}
The effective electric field is defined by
\begin{align}
    {{\bf{e}}_j} &= - {\partial _t}{{\bf{a}}_j} - \nabla {a_{0j}} 
    \nonumber \\
   &=  - 2\hbar \sigma  \cdot \left( {\partial _t}{{\bf{d}}_j} 
     \times \nabla {{\bf{d}}_j} \right).
\label{eq_e}
\end{align}
If we define a staggered potential by
\be
{V_{st}} = {\left( { - 1} \right)^j}{\left[ {{a_{0j}}} 
    \right]_{ \uparrow  \uparrow }} 
=  - {\left( { - 1} \right)^j}{\left[ {{a_{0j}}} 
    \right]_{ \downarrow  \downarrow }},
\label{eq_Vst}
\ee
then $V_{st}$ is a smooth function with respect to $z$,
as shown in Fig.~\ref{fig_Vst}.
In the case of ferromagnets,
the effective electric field given by Eq. (\ref{eq_e}) leads to the spin motive force.
However, in the case of antiferromagnets, 
it is not clear whether the field given by Eq. (\ref{eq_e}) leads to the spin motive force.
Nevertheless, as will be shown later,
the field (\ref{eq_e}) leads to the spin motive force 
in the case of antiferromagnets as well,
and that is confirmed by numerical simulation.
  \begin{figure*}[float]
    \begin{center}
      \includegraphics[width=0.8 \linewidth]{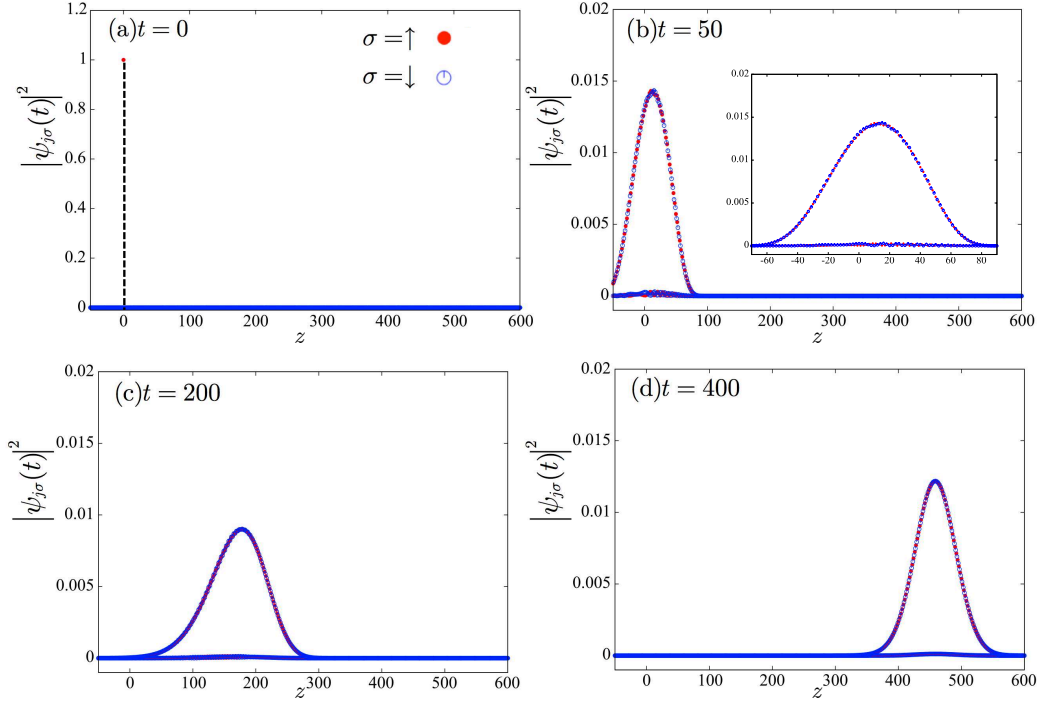}
    \end{center}
    \caption{ \label{fig_te}
     (Color online)
      Time evolution of the wave packet.
      The number of lattice sites is 600.
      We take $\ell=200$ and $J/\eta=5$.
      For $T$, we take $\eta T/\hbar = 5$.
      As the initial state at $t=0$, we put a spin-up state
      at $j=0$ (a).
      (The dashed line is a visual guide.)
      The wave packet quickly becomes a Gaussian form consisting of
      spin-up and spin-down states (b).
      Inset of (b) shows a magnified view of the wave packet
      near the origin.
      We clearly observe the motion of the wave packet 
      in the positive $z$-direction [(c) and (d)].
      The unit of time is taken as $\hbar/\eta$.
    }
  \end{figure*}

To verify the creation of the spin motive force in the system,
we solve the time dependent Schr\"{o}dinger equation for the conduction
electron with the Hamiltonian Eq. (\ref{eq_H}).
We represent the eigenstate of the system by 
$\left| {\psi \left( t \right)} \right\rangle$.
The wave function ${\psi _{j\sigma }}\left( t \right)$
of the conduction electron at site $j$ with spin $\sigma$
is defined by:
\be
\left| {\psi \left( t \right)} \right\rangle  
= \sum\limits_{j,\,\,\sigma } {{\psi _{j\sigma }}\left( t \right)
  c_{j\sigma }^\dag \left| 0 \right\rangle }.
\ee
Here, $\left| 0 \right\rangle$ denotes the vacuum state.
We solve the time evolution of ${\psi _{j\sigma }}\left( t \right)$
using the fourth-order Runge-Kutta method
under an open boundary condition.
The time evolution of the wave function is shown in Fig.~\ref{fig_te}
in the case of $J/\eta=5$.
As an initial condition at $t=0$, we take the spin-up state at $j=0$
in the form of a static Gaussian wave packet,
${\psi _{j\sigma }}\left( 0 \right) 
= C\delta_{\sigma,\uparrow}
\exp \left( {- {{j^2}}/{{\lambda ^2}}} \right)$
with $C$ as the normalization constant [Fig.~\ref{fig_te}(a)].
We consider a localized state at the origin and take $\lambda=0.5$.
The initial localized spin-up state quickly 
expands, as shown in Fig.~\ref{fig_te}(b).
The wave packet consists of spin-up states at A sublattice
and spin-down states at B sublattice.
This feature is understood from the potential term mentioned above.
As time elapses, 
the wave packet clearly moves to the direction of 
the positive $z$-axis [Figs.~\ref{fig_te}(c) and \ref{fig_te}(d)].
This motion of the wave packet demonstrates the presence 
of the spin motive force in the antiferromagnet.
We carried out a numerical simulation 
for $J/\eta=3$ and obtained a similar result (not shown).

We note that the direction of the wave function propagation
of the conduction electron depends on the external magnetic field.
If we reverse the direction of the magnetic field,
the direction of the wave function propagation is reversed.
If we change the sign of $J$, which affects the direction,
the direction of the wave function propagation is reversed as well.
We also note that the initial spin state does not affect
the propagation direction.
This is because the spin state of the conduction electron
takes the lower energy state upon propagation
due to the potential energy associated with 
the exchange interaction.

\begin{figure}
    \begin{center}
      \includegraphics[width=0.8 \linewidth]{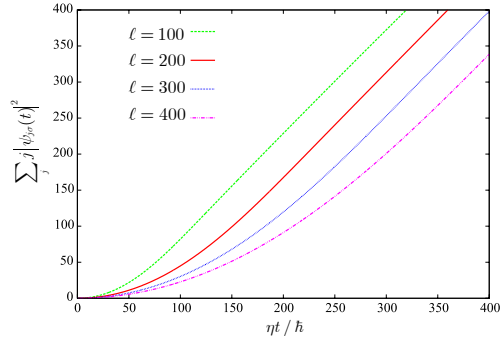}
    \end{center}
    \caption{ \label{fig_cm}
     (Color online)
      Time evolution of the expectation value of 
      the position of the conduction electron
      for $\ell = 100, 200, 300 and 400$.
      The time dependence is well described by a particle motion
      under a uniform field when the wave packet is far from the edge
      of the domain wall (see the text.)
      The time dependence changes around $t\sim \ell$
      when the wave packet reaches the edge of the domain wall.
    }
  \end{figure}

The time dependence of the wave packet motion becomes
more clearly seen from
the expectation value of the position of the conduction electron,
\be
\sum\limits_j {{j}{{\left| {{\psi _{j\sigma }}
        \left( t \right)} \right|}^2}}.
\ee
This quantity is shown in Fig.~\ref{fig_cm}.
When the wave packet is far from the edges of the domain wall,
the motion of the wave packet is in agreement with
the motion of the corresponding classical particle
under a constant acceleration.
We find
\be
\sum\limits_j {{j}\left| {{\psi _{j\sigma }}\left( t \right)} \right|^2} 
\simeq \frac{{{\pi ^2}}}{{4m\ell T}}{t^2},
\ee
for $t\ll 2\ell$.
This is understood from the force acting on the conduction electron
created by 
the effective electric field Eq. (\ref{eq_e}).
From the fitting of the data, we find that $m\simeq 0.5$, 
which is consistent with the picture above.

We carried out a similar calculation in the ferromagnetic domain wall
case and found the same result.
Within the strong coupling limit, there is no difference.

The appearance of the spin motive force in the precessing 
antiferromagnetic domain wall is clearly understood as follows.
The key is the staggered potential $V_{st}$.
A naive expectation is that the spin motive force cancels out
because of the sign difference in the spin motive force
Eq.~(\ref{eq_Vst})
at A and B sublattices.
However, the spin of the conduction electron flips at each hopping.
The spin state of the conduction electron changes at each hopping 
through the term $U_j^\dag {U_{j + 1}}i{\sigma _y}$.
We note that $U_j^\dag {U_{j + 1}}$ is close to
the unit matrix because of the smooth variation of 
the staggered magnetization with respect to $j$ and $t$.
Because of the presence of the term $i{\sigma_y}$,
the spin of the conduction electron flips.
The spin flip leads to the sign change in $V_{st}$,
while there is a sign difference in $V_{st}$ between 
A and B sublattices.
Therefore, there are two sign changes,
and thus no cancellation occurs for the spin motive force
created by the staggered potential $V_{st}$.
We note that the gauge field appearing in the components of 
the matrix $U_j^\dag {U_{j + 1}}$
does not play an important role.
In a one-dimensional antiferromagnetic domain wall case,
the effect of the gauge field in $U_j^\dag {U_{j + 1}}$ vanishes
under the precession created by the magnetic field.
\tm{
The vector potential given by Eq. (\ref{vec_po}) does not play an important role, which is clearly understood as follows.
In the one-dimensional antiferromagnetic domain wall, which is shown in Fig. \ref{fig_afdw}, the $z$-component of the vector potential\cite{Tserkovnyak2008} is written as
\begin{align*}
  [{{\bf a}}_j]_z =\begin{cases}
  -\hbar \sigma_z (\nabla_j \phi)\sin^2\displaystyle\frac{\theta_j}{2} &(j\in A)\\
\\
  -\hbar (-i\sigma_y) \left( \sigma_z (\nabla_j \phi)\sin^2\displaystyle\frac{\theta_j}{2} \right) (i\sigma_y) &(j\in B)
\end{cases}.
\end{align*}
We note that the azimuthal angle $\phi_j$ is independent of site $j$.
Therefore, Eq. (4) is negligible.
}

Now, we comment on the semiclassical approximation.\cite{Cheng2012,Cheng2014}
One can derive an effective Hamiltonian by
starting from the band picture and then including
the spatial variation of the spins.
In such an approximation, there appears a term 
that depends on the dynamics and spatial variation of the spins.
However, the term vanishes in the strong coupling limit.
This is simply because the semiclassical approximation
is not justified in the strong coupling limit
because the spatial variation of the spins
is not properly taken into account.
In contrast, the spatial variation of the spins
is exactly taken into account in our formulation
based on the unitary transformation.

To summarize, we have numerically demonstrated 
the creation of the spin motive force in a rigidly
precessing antiferromagnetic domain wall.
We have solved the time dependent Sch\"{o}dinger equation
and clearly shown the motion of the wave packet of 
the conduction electron
that reflects the presence of the spin motive force.
The motion of the conduction electron is 
understood by the combined effect 
of the force created by the 
staggered potential $V_{st}$
and the spin flip at each hopping.
Our numerical simulation suggests that 
there is no qualitative distinction between
antiferromagnets and ferromagnets 
in the strong coupling limit
with respect to the spin motive force.

{\footnotesize \section*{Acknowledgements} 
We thank K. Kubo and A. Shitade for helpful discussions.

\bibliography{../../../references/tm_library2}

\end{document}
%